\overfullrule=0pt
\input harvmac
\def\a{{\alpha}}

\def\l{{\lambda}}
\def\b{{\beta}}
\def\g{{\gamma}}
\def\d{{\delta}}
\def\e{{\epsilon}}

\def\half{{1\over 2}}
\def\p{{\partial}}

\def\t{{\theta}}

\Title{\vbox{\hbox{IFT-P.035/2001 }}}
{\vbox{
\centerline{\bf Relating the RNS and Pure Spinor
Formalisms for the Superstring}}}
\bigskip\centerline{Nathan Berkovits\foot{e-mail: nberkovi@ift.unesp.br}}
\bigskip
\centerline{\it Instituto de F\'\i sica Te\'orica, Universidade Estadual
Paulista}
\centerline{\it Rua Pamplona 145, 01405-900, S\~ao Paulo, SP, Brasil}

\vskip .3in
Recently, the superstring was covariantly quantized using the 
BRST-like operator $Q = \oint \lambda^\alpha d_\alpha$ where
$\lambda^\alpha$ is a pure spinor and $d_\alpha$ are the fermionic
Green-Schwarz constraints. By performing a field redefinition and
a similarity transformation, this BRST-like operator is mapped to the sum
of the Ramond-Neveu-Schwarz BRST operator and $\eta_0$ ghost.
This map is then used to relate physical
vertex operators and tree amplitudes in the two formalisms.
Furthermore, the map implies the existence of a $b$ ghost in the
pure spinor formalism which might be useful for loop amplitude computations.

\vskip .3in
\centerline{\it Dedicated to my father on his 75th birthday}

\Date {April 2001}

\newsec{Introduction}

Super-Poincar\'e covariant quantization of the ten-dimensional
superstring is an important problem which has attracted many
different approaches.
Recently, a new approach 
\ref\superp{N. Berkovits, {\it Super-Poincar\'e Covariant Quantization of the
Superstring}, JHEP 04 (2000) 018, hep-th/0001035.}
was proposed using a BRST-like operator
$Q=\oint\l^\a d_\a$ where $\l^\a$ is a pure spinor worldsheet variable
and $d_\a$ are the super-Poincar\'e covariant
fermionic Green-Schwarz constraints.
Unlike all previous approaches, this pure spinor approach was
used to construct physical vertex operators
and compute non-vanishing superstring tree amplitudes in
a manifestly super-Poincar\'e covariant manner.

Because the BRST-like operator $Q$ involves second-class constraints,
its construction is non-conventional so the validity of the pure
spinor formalism needs to be verified. Previous steps in this direction
include showing that the cohomology of $Q$ reproduces the
light-cone Green-Schwarz (GS) spectrum \ref\cohom
{N. Berkovits, {\it Cohomology in the Pure Spinor Formalism for the
Superstring}, JHEP 09 (2000) 046, hep-th/0006003.}
and that tree amplitudes involving
external massless states (with up to four fermions)
coincide with the Ramond-Neveu-Schwarz (RNS)
computations \ref\val{N. Berkovits and B.C. Vallilo, 
{\it Consistency of Super-Poincar\'e Covariant
Superstring Tree Amplitudes}, JHEP 07 (2000) 015, hep-th/0004171.}.

In this paper, we shall provide further evidence for the validity of the
pure spinor formalism by 
finding a field redefinition and similarity 
transformation which maps the BRST-like operator $Q=\oint \l^\a d_\a$
into $Q' = Q_{RNS} +\oint\eta$
where $Q_{RNS}$ is the RNS BRST operator and $\eta$ is the RNS
variable coming from
fermionization of the super-reparameterization $(\b,\g)$ ghosts
\ref\fms{D. Friedan, E. Martinec and S. Shenker, {\it
Conformal Invariance, Supersymmetry and String Theory},
Nucl. Phys. B271 (1986) 93.}.
This map will then
be used to relate physical vertex operators and tree
amplitudes in the two formalisms.

In a fixed picture, the cohomology of $Q'=Q_{RNS}+\oint \eta$
in the ``large'' RNS Hilbert space is the same as the cohomology of
$Q_{RNS}$ in the ``small'' RNS Hilbert space.\foot
{In the language of \fms, the ``large'' and ``small'' RNS Hilbert
spaces refer to the spaces with and without the $\xi$ zero mode.}
However, as will be 
shown, picture-changing is a gauge transformation if one
uses $Q'$
in the ``large'' Hilbert space. So although any physical state
can be represented by vertex operators
in different pictures in the cohomology
of $Q_{RNS}$, all such vertex operators are equivalent
in the cohomology of $Q'$. This will be important since
spacetime-supersymmetry transformations in the RNS formalism only
close up to picture-changing.

As was shown in \ref\ufive{N. Berkovits, {\it
Quantization of the Superstring with Manifest
U(5) Super-Poincar\'e Invariance}, Phys. Lett. B457 (1999) 94, 
hep-th/9902099.},
there is a field redefinition
from the ten-dimensional RNS variables $(x^m,\psi^m,b,c,\beta,\gamma)$
into GS-like variables which
manifestly preserves six spacetime supersymmetries and a $U(5)$ subgroup
of the Wick-rotated SO(10) Lorentz group. The worldsheet variables
in this GS-like description of the superstring consist
of ten spacetime coordinates $x^m$ for $m=0$ to 9, six
fermionic superspace coordinates $(\t^+, \t^a)$
and their conjugate momenta $(p_+, p_a)$ for $a=1$ to 5,
and two chiral bosons $(s,t)$. 

The next step in relating the RNS and pure spinor formalisms is to
add a ``topological'' sector to the GS-like variables consisting of
ten fermionic superspace coordinates and their conjugate momenta,
$(\t_{ab},p^{ab})$, as well as ten bosonic coordinates and their
conjugate momenta, $(u_{ab}, v^{ab})$. In order that physical
states are unaffected by this topological sector, the BRST operator
$Q'$ will be modified to $Q_{U(5)} = Q' +\half\oint
u_{ab} p^{ab}+ ... .$ Because of the standard quartet mechanism, states
in the cohomology of $Q_{U(5)}$ will be independent of the $(\t_{ab},p^{ab},
u_{ab},v^{ab})$ variables.

The final step in relating the two formalisms is to find a similarity
transformation $R$ such that $e^{-R} Q_{U(5)} e^{R} = \oint \l^\a d_\a$
where $(\t^a,\t^+)$ combines with $\t_{ab}$ to form a covariant 
spinor $\t^\a$ for $\a=1$ to $16$, 
$(p_a,p_+)$ combines with $p^{ab}$ to form a covariant 
spinor $p_\a$, and $d_\a$ is constructed from $(x^m,p_\a,\t^\a)$
in the usual super-Poincar\'e covariant manner. Furthermore,
the chiral bosons $(s,t)$ of the $U(5)$ formalism combine with
the bosons $(u_{ab},v^{ab})$ of the topological sector to form
a pure spinor $\l^\a$ and its conjugate momentum where $u_{ab}$
parameterizes the ten-dimensional complex coset $SO(10)/U(5)$.

So after using the field redefinition to
write any physical RNS vertex operator $U_{RNS}$ in terms
of $U(5)$ variables, one can construct a vertex operator
$U$ in the cohomology of $Q=\oint\l^\a d_\a$ by defining $U=e^{-R} U_{RNS}
e^{R}$. 
Using this map from physical RNS vertex operators
to physical pure spinor vertex operators, it is straightforward
to show that the tree amplitudes in the two formalisms are identical.
This justifies the rather unconventional normalization prescription
of \superp\ for integrating over worldsheet zero modes in the pure
spinor formalism.

The only subtlety in relating the two formalisms
is that although the operators $Q_{U(5)}$ and $R$ are
manifestly spacetime supersymmetric, they are not
manifestly Lorentz invariant. So
pure spinor vertex operators $U$ obtained by this map
do not necessarily
transform covariantly under Lorentz transformations. 
Note that the similarity transformation guarantees only that
$U$ transforms covariantly up to a BRST-trivial operator.
However, evidence will be given that, with one exception which will
be discussed in the following paragraph,
there is always a suitable gauge choice
for $U$ which transforms covariantly.

The one exception is the ghost-number $-1$ operator $\int d^2 z ~\mu(z)~b(z)$
where $b$ is the RNS $b$ ghost and
$\mu(z)$ is a Beltrami differential. When mapped to the pure spinor
formalism, there is no gauge choice for
which this operator is super-Poincar\'e
invariant. Nevertheless, it will be argued that the OPE of this operator
with any physical operator of positive ghost number can be written
in super-Poincar\'e covariant form. 
Furthermore, this operator can be used in the pure spinor formalism
to construct integrated vertex operators from unintegrated operators,
to relate string antifields and fields, and to define tree amplitudes
in a worldsheet reparameterization invariant manner.
It is hoped that this ghost-number $-1$
operator will also
be useful for computing loop amplitudes using the pure spinor formalism.

Section 2 of this paper will review the pure spinor formalism of
the superstring, and section 3 will review the RNS formalism using
a modified definition of physical states in which picture
changing is a gauge transformation. In section 4, a map will be
found which takes the sum of the RNS BRST operator and $\eta_0$
ghost into the BRST-like operator in the pure spinor formalism.
In section 5, this map will be used to relate physical
vertex operators and tree amplitudes in the two formalisms.
In section 6, the $b$ ghost will be constructed in the pure
spinor formalism and 
section 7 will conclude with open problems and speculations.

\newsec{Review of the Pure Spinor Formalism}

\subsec{Pure spinors}

In this section, we shall review the relevant features of the
pure spinor formalism of the superstring. Following the work
of Siegel  
\ref\siegel
{W. Siegel, {\it Classical Superstring Mechanics}, Nucl. Phys. B263 (1986)
93.}, the action in this formalism is constructed using
a first-order action for the $\t^\a$ worldsheet variables where
the fermionic conjugate momenta, $p_\a$, are independent worldsheet
variables. In addition to the worldsheet variables $(x^m,
\t^\a, p_\a)$ for $m=0$ to 9 and $\a=1$ to $16$, the action also depends 
on a bosonic ``ghost" variable $\l^\a$ satisfying the pure spinor
condition
\eqn\pure{\l^\a \g^m_{\a\b} \l^\b =0 {\rm ~~~for~~} m=0{~to~}9}
where $\g^m_{\a\b}$ and $\g^{m~\a\b}$
are $16\times 16$ symmetric matrices which
form the off-diagonal blocks of the $32\times 32$ ten-dimensional
$\Gamma$-matrices
in the Weyl representation.

In worldsheet conformal gauge, the worldsheet action is 
\eqn\action{S = \int d^2 z [\half \p x^m \overline\p x_m + p_\a \overline
\p\t^\a]
+ S_\l}
with the free-field OPE's
\eqn\freeop{x^m(y) x^n(z) \to -\eta^{mn}\log|y-z|^2,\quad p_\a(y)\t^\b(z)
\to (y-z)^{-1} \d_\a^\b,}
where $S_\l$ is the action for the pure spinor variable which will
be described more explicitly in the following paragraphs.
In the above action and in the rest of this paper, we shall
ignore the right-moving degrees of freedom. The results of this
paper are easily generalized to describe the heterotic, Type I,
or Type II superstrings by choosing the right-moving sector
appropriately.

Since $\l^\a$ is constrained by \pure, it is convenient to solve
this constraint when defining $S_\l$. A
parameterization of $\l^\a$
which preserves a $U(5)$ subgroup of (Wick-rotated)
$SO(10)$ is\foot{To simplify
comparison with the $U(5)$ formalism of \ufive, the $u_{ab}$
variables defined here differ from those of \superp\ by a
factor of $e^s$ (which was called $\gamma$ in \superp).}
\eqn\paramone{\l^+ = e^s ,\quad \l_{ab} =  u_{ab},\quad
\l^a = -{1\over 8} e^{-s}\e^{abcde} u_{bc} u_{de}}
where $a=1$ to 5,
$u_{ab}= - u_{ba}$ are ten complex variables parameterizing
the $SO(10)/U(5)$ coset\foot{
If one does not Wick rotate, the $u_{ab}$ variables of \paramone\
parameterize the compact space $SO(9,1)/(U(4)\times {\cal R}^9)$
where $U(4)$ is a subgroup of the transverse $SO(8)$ rotations
and ${\cal R}^9$ represents the nine light-like boosts generated
by $M^{+m}$ \ref\howe{P.S. Howe, {\it Pure Spinor Lines in Superspace and
Ten-Dimensional Supersymmetric Theories}, Phys. Lett. B258 (1991) 141.}.},
$s$ is a complex phase, and the SO(10)
spinor $\l^\a$ has been written in terms of its irreducible $U(5)$
components which transform as $(1_{{5\over 2}}, \overline{10}_{\half},
5_{-{3\over 2}})$ representations of $SU(5)_{U(1)}$.\foot{A simple
way to obtain these $U(5)$ representations is to write an $SO(10)$
spinor using $[\pm\pm\pm\pm\pm]$ notation 
where Weyl/anti-Weyl spinors have an odd/even number of $+$ signs.
The $1_{5\over 2}$ component of $\l^\a$ is the component with five
$+$ signs, the 
$\overline{10}_{\half}$ component has three $+$ signs, and the 
$5_{-{3\over 2}}$ component has one $+$ sign.}
The $\l^\a$
parameterization of \paramone\ is possible whenever $\l^+\neq 0$.

Using the above parameterization of $\l^\a$, one can define
\eqn\actone{S_\l = \int d^2 z [\overline\p t \p s  - \half
v^{ab}\overline\p u_{ab}]} 
where $t$ and $v^{ab}$ are the conjugate momenta to $s$ and $u_{ab}$
satisfying the OPE's
\eqn\opel{t(y)~ s(z) \to \log (y-z),\quad v^{ab} (y) ~u_{cd}(z) \to
\d_c^{[a} \d_d^{b]} (y-z)^{-1}.} 
Note that the factor of $\half$ in the $v^{ab} 
\overline\p u_{ab}$ term has been
introduced to cancel the factor of 2 from $u_{ab}=
-u_{ba}$.
Also note that $s$ and $t$ are chiral bosons, so their
contribution to \actone\ needs to be supplemented by a chirality constraint.
Furthermore, the zero modes of $s$ and $t$ can only appear through
the exponentials $e^{ms+nt}$ for integers $m$ and $n$.

One can construct $SO(10)$
Lorentz currents $N^{mn}$ out of these free
variables as\foot{This construction
can be obtained from \superp\ by replacing $u_{ab}$ with $u_{ab} e^s$
as explained in footnote 3. Similar
free-field constructions can be found in \ref\bars{I. Bars,
{\it Free Fields and New Cosets of Current Algebras}, Phys. Lett. B255
(1991) 353.}.}
\eqn\lorentz{N = {1\over{\sqrt{5}}}( {1\over 4}
u_{ab} v^{ab} + {5\over 2}\p t - {5\over 2}\p s), \quad
N_a^b =  u_{ac} v^{bc} -{1\over 5}\d_a^b u_{cd}v^{cd}, }
$$N^{ab} = e^s v^{ab}, \quad
N_{ab} =  e^{-s}(2 \p u_{ab} - u_{ab}\p t -2 u_{ab}\p s
+u_{ac} u_{bd} v^{cd} -\half 
u_{ab} u_{cd} v^{cd}) $$
where $N^{mn}$ has been written in terms of its $U(5)$ components
$(N,N_a^b,N^{ab},N_{ab})$ 
which transform as $(1_0,24_0,10_2,\overline{10}_{-2})$ representations
of $SU(5)_{U(1)}$.
The Lorentz currents of \lorentz\ can be checked to satisfy the OPE's
\eqn\Nlope{N^{mn}(y) \l^\a (z) \to 
\half(\g^{mn})^\a{}_\b {{\l^\b(z)}\over{y-z}},}
\eqn\nope{N^{kl}(y) N^{mn}(z) \to
{{\eta^{m[l} N^{k]n}(z) -
\eta^{n[l} N^{k]m}(z) }\over {y-z}} - 3
{{\eta^{kn} \eta^{lm} -
\eta^{km} \eta^{ln}}\over{(y-z)^2}}  .}
So although $S_\l$ is not manifestly Lorentz covariant, any OPE's
of $\l^\a$ and $N^{mn}$ which are
computed using this action are manifestly covariant.

In terms of the free fields, the stress tensor is 
\eqn\stress{T = -\half \p x^m \p x_m - p_\a \p\t^\a + \half v^{ab}\p u_{ab}
+ \p t\p s + \p^2 s}
where the $\p^2 s$ term is included so that the Lorentz currents of
\lorentz\ are primary fields. This stress tensor has zero central charge
and can be written in manifestly Lorentz invariant notation as\ref\nbersh
{N. Berkovits and M. Bershadsky, unpublished.}
\eqn\covstress{T = 
 -\half \p x^m \p x_m - p_\a \p\t^\a + {1\over {10} } N_{mn} N^{mn}
- \half (\p h)^2 -2 \p^2 h}
where $h$ is a Lorentz scalar defined in terms of the free fields
by 
\eqn\hdef{\p h = {1\over 4} u_{ab} v^{ab} + \half\p t + {3\over 2} \p s.}
Note that
$h$ has no singularities with $N^{mn}$ and satisfies the OPE's
$$h(y) h(z) \to - \log(y-z),\quad 
\p h (y) \l^\a (z) \to \half (y-z)^{-1} \l^\a(z).$$ 
The operator $2\oint \p h$ will be identified with the
ghost-number operator so that $\l^\a$ carries ghost number $+1$.

\subsec{Physical states}

Physical states in the pure spinor formalism are defined as
super-Poincar\'e covariant
ghost-number $+1$ states in the cohomology of the BRST-like operator
\eqn\brst{Q = \oint \l^\a d_\a}
where 
\eqn\ddef{d_\a= p_\a -\half\g_{\a\b}^m \t^\b \p x_m -{1\over 8}
\g_{\a\b}^m \g_{m~\g\d}\t^\b\t^\g\p\t^\d}
is the super-Poincar\'e covariant Green-Schwarz constraint\siegel.
Note that $Q^2=0$ since $d_\a$ satisfies the OPE's
\eqn\oped{d_\a(y) d_\b(z) \to -(y-z)^{-1} \g_{\a\b}^m \Pi_m(z),\quad 
d_\a(y) \Pi^m(z) \to  (y-z)^{-1} \g_{\a\b}^m \p\t^\b(z),}
where $\Pi^m = \p x^m +\half\t \g^m \p\t$
is the supersymmetric momentum. Furthermore, $Q$ is spacetime
supersymmetric since $d_\a$ anticommutes with the spacetime supersymmetry
generators\siegel
\eqn\defqq{
q_\a= \oint (p_\a +\half\g_{\a\b}^m \t^\b \p x_m +{1\over {24}}
\g_{\a\b}^m \g_{m~\g\d}\t^\b\t^\g\p\t^\d).}

For $Q$ to be hermitian, $\l^\a$
must be defined to be a hermitian operator. Although the pure spinor
condition of \pure\ has no real non-vanishing solutions, this does not
cause any inconsistency. Since the Hilbert space
inner product does not have a positive
definite norm, there is no reason why $(\l^\a)^\dagger \l^\a$ must
be an operator with positive eigenvalues
\ref\pol
{J. Polchinski, private communication.}. Note, however, that when
$\l^\a$ is hermitian, the
$(s,t,u_{ab},v^{ab})$ variables have strange hermiticity properties.
For example, using the notation of footnote 5,
$(\l^{+++++})^\dagger=\l^{+----}$
implies that $(e^s)^\dagger = -{1\over 8} e^{-s} \e^{abcd5} u_{ab} u_{cd}.$

The physical vertex operator for the massless super-Maxwell
multiplet is $U= \l^\a A_\a(x,\t)$ where $A_\a(x,\t)$ is the
spinor gauge potential of super-Maxwell theory. $QU=0$ and 
$\d U= Q \Lambda$ implies that
$D_\a (\g_{mnpqr})^{\a\b} A_\b =0$ and $\d A_\a = D_\a \Lambda$
where $D_\a = {\p\over{\p\t^\a}} +\half \t^\b \g^m_{\a\b} \p_m$ and
${mnpqr}$ is an arbitrary five-form direction \howe.
These are the superspace
equations of motion and gauge invariances of super-Maxwell theory.

Using the $\l^\a$ parameterization defined in \paramone,
pure spinor vertex operators can be described in terms of unconstrained
variables. However, not every function of the unconstrained variables
is an allowed pure spinor
vertex operator. The requirement of super-Poincar\'e covariance
implies that the function
must transform
as a finite dimensional representation under commutation with the
Lorentz generators of \lorentz, which implies that the $(s,t,u_{ab},v^{ab})$
variables can only appear in the Lorentz covariant
combinations $(\l^\a,N_{mn},\p h)$ of \paramone, \lorentz, and
\hdef.

\subsec{Tree amplitudes}

To compute $N$-point tree amplitudes, one needs 3 dimension-zero
vertex operators $U$ and $N-3$ dimension-one vertex operators $V$ which
will be integrated over the real line. For a given physical state described
by the dimension-zero operator $U$, the dimension-one operator $V$ can
be defined by requiring that $QV=\p U$. For the super-Maxwell vertex
operator, one can check that 
\eqn\supermax{V = \p \t^\a A_\a (x,\t) + 
\Pi^m B_m (x,\t) + d_\a W^\a (x,\t) +\half N^{mn} F_{mn}(x,\t)}
where $B_m = {1\over 8}D_\a \g_m^{\a\b} A_\b$,
$W^\b = {1\over{10}}\g_m^{\a\b} (  D_\a B^m -\p^m A_\a)$, and $F_{mn} =
{1\over 8}D_\a (\g_{mn})^\a{}_\b W^\b = \p_{[m} B_{n]}$.

Tree amplitudes are then defined by the correlation function
\eqn\ampl{A = \int dz_4 ... dz_N \langle ~U_1(z_1) ~ U_2(z_2) ~
U_3(z_3) ~\int dz_4 V_4(z_4) ... \int dz_N V_N(z_N) ~\rangle.}
As shown in \val, 
this definition is independent of which three external
states are represented by unintegrated vertex operators.
Since the action of \action\ is quadratic, the free field OPE's can be
used to perform the integration over the non-zero modes of the
worldsheet fields. The resulting expression,
$$A = \int dz_4 ... dz_N \langle ~ f(k_r,\eta_r,z_r,\l,\t)~\rangle $$
\eqn\result{
 = \int dz_4 ... dz_N \langle ~\l^\a\l^\b \l^\g f_{\a\b\g}
(k_r,\eta_r,z_r,\t)~\rangle,}
only depends on the external momenta $k_r$ and polarizations $\eta_r$, 
and on the zero
modes of the $\l^\a$ and $\t^\a$ fields. Furthermore, the
expression is cubic in $\l^\a$
since $U$ carries ghost-number $+1$ and $V$ carries ghost-number $0$.

The $(\l,\t)$ zero-mode integration will be defined by
\eqn\zeromode{
\langle ~\l^\a\l^\b \l^\g f_{\a\b\g}
(k_r,\eta_r,z_r,\t)~\rangle }
$$= ({\p\over{\p\t}} \g_{mnp}
{\p\over{\p\t}})
({\p\over{\p\t}}\g^m)^\a
({\p\over{\p\t}}\g^n)^\b
({\p\over{\p\t}}\g^p)^\g f_{\a\b\g}
(k_r,\eta_r,z_r,\t)|_{\t=0}.$$
In other words, only the term proportional to
$(\t\g^{mnp}\t)(\g_m\t)_\a (\g_n\t)_\b(\g_p\t)_\g$ in
$f_{\a\b\g}$ contributes to the scattering amplitude.
Since $(\t\g^{mnp}\t)(\l\g_m\t)(\l\g_n\t)(\l\g_p\t)$ does not equal
$Q\Lambda$ for any Lorentz covariant $\Lambda$, 
the zero mode prescription of \zeromode\ implies that
\eqn\ampgau{\langle \l^\a\l^\b \l^\g f_{\a\b\g}+ Q\Lambda\rangle
=\langle \l^\a\l^\b \l^\g f_{\a\b\g}\rangle,}
so the amplitude is gauge invariant.  
This tree amplitude prescription was shown in \superp\
to be spacetime supersymmetric and was shown by explicit computation
in \val\ to coincide with the RNS tree
amplitude prescription for massless external states with up to four fermions,
In subsection (5.4) of this paper, the map from pure spinor vertex operators
to RNS vertex operators will be used to argue
that this tree amplitude prescription agrees with
the RNS prescription for arbitrary external states. 

\newsec{RNS Picture-Changing as a Gauge Transformation}

In this section, we shall discuss the relevant features of the RNS
formalism using a modified 
definition of physical states where picture-changing
is a gauge transformation.

\subsec{Picture-changing}

Physical vertex operators in the RNS formalism can be defined as
ghost-number 
$+1$ operators\foot{
As in \ref\sft{N. Berkovits, {\it Super-Poincar\'e Invariant
Superstring Field Theory}, Nucl. Phys. B450 (1995) 90, hep-th/9503099.},
we shall define the ghost-number
operator as $\oint(cb+\eta\xi)$ so that $\eta$ carries ghost-number $+1$.
This definition agrees with the ghost-number operator $\oint(cb - \p\phi)$
of \fms\
at zero picture, but has the advantage of commuting with picture-changing
and spacetime-supersymmetry.} 
$U$ which satisfy \ref\top{N. Berkovits and C. Vafa,
{\it N=4 Topological Strings}, 
Nucl. Phys. B433 (1995) 123, hep-th/9407190.}
\eqn\phys{\eta_0 U = Q_{RNS} U=0, \quad 
\d U = Q_{RNS} \eta_0 \Lambda}
where 
\eqn\brstr{Q_{RNS}=\oint[c(-\half \p x^m\p x_m -\half\psi^m \p\psi_m
-\eta\p\xi-\half\p\phi\p\phi -\p^2\phi-b\p c)}
$$ + \eta e^\phi \psi^m\p x_m
-\eta\p\eta e^{2\phi} b]$$
is the RNS BRST operator and
$\eta_0$ is the zero mode coming from fermionizing 
the $(\b,\g)$ super-reparameterization ghosts 
as $\b=\p\xi e^{-\phi}$ and $\g = \eta e^\phi$ \fms.
Note that $\eta_0 U =0$ implies that
$U$ is in the ``small" Hilbert space and that any gauge parameter in
the ``small'' Hilbert space, $\Omega$, can be written as
$\Omega= \eta_0 \Lambda$ for some $\Lambda$. 

It might seem
surprising that $Q_{RNS}$ and $\eta_0$ appear symmetrically in the definition
of \phys. However, note that in the ``large'' Hilbert space, both
$Q_{RNS}$ and 
$\eta_0$ have trivial cohomologies. The cohomology of $Q_{RNS}$ is
trivial in the ``large'' Hilbert space since any state $U$ satisfying
$Q_{RNS}U=0$ can be written as $U= Q_{RNS}(c \xi\p\xi e^{-2\phi} U)$.
And the cohomology of $\eta_0$
is trivial in the ``large'' Hilbert space since any state $U$ satisfying
$\eta_0 U=0$ can be written as $U= \eta_0 ( \xi U)$.

Using the picture-raising operation,
\eqn\raising{U_{P+1} = \{Q_{RNS},\xi U_P\} }
or the picture-lowering operation, 
\eqn\lowering{U_{P-1} 
= \{\eta_0 , c\xi\p\xi e^{-2\phi} U_P\}, }
any physical state represented by the vertex operator $U_P$
of picture $P$ can also be represented by a vertex operator
of arbitrarily higher or lower picture.
As will now be shown, this redundancy in describing physical states
can be removed by defining physical vertex operators as ghost-number
$+1$ operators in the cohomology of $Q' = Q_{RNS}+\eta_0$.

\subsec{Cohomology of $Q'$}

To show that the cohomology of $Q'=Q_{RNS}+\eta_0$ correctly reproduces
the physical spectrum, consider a vertex operator $U$ annihilated by
$Q'$ of the form $U = \sum_{P=P_-}^{P_+} U_P$ where $U_P$ carries
picture $P$ and $P_\pm$ are the highest/lowest picture.\foot{Only
vertex operators involving a finite number of picture will be allowed
in the Hilbert space. If vertex operators involving arbitrarily high
(or low) picture were allowed in the Hilbert space, the cohomology
of $Q'$ would be trivial.} Since $Q_{RNS}$ carries zero picture and
$\eta_0$ carries $-1$ picture, $Q' U=0$ implies that
$\eta_0 U_{P_-}=0$, which implies
that $U_{P_-}= \eta_0 \Omega_{P_- +1}$ for some 
$\Omega_{P_- +1}$ of picture $P_- +1$. 
Then after the gauge transformation $\d U = -Q' \Omega_{P_- +1}$, 
$U = \sum_{P = P_- +1}^{P_+} U_P$ where $U_{P_- +1}$ is now
$U_{P_- +1} - Q_{RNS}\Omega_{P_- +1}$. This procedure can be continued
until $U= U_{P_+}$ where $Q' U_{P_+}=0$. But this implies that
$Q_{RNS} U_{P_+} = \eta_0 U_{P_+} =0$, so $U_{P_+}$ is a physical state
using the definition of \phys.

To show that this physical state is not pure gauge,
suppose that $U = U_{P_+} = Q' \Lambda$ for some
$\Lambda$. Then $\Lambda = \Lambda_{P_+} + \Lambda_{P_+ +1}$
where 
\eqn\whg{U= Q_{RNS}\Lambda_{P_+} + \eta_0\Lambda_{P_+ +1},}
\eqn\whh{Q_{RNS}\Lambda_{P_+ +1} = \eta_0\Lambda_{P_+} =0.}
The
$Q_{RNS}$ and $\eta_0$ cohomologies are trivial in the ``large''
Hilbert space so 
\whh\ implies that 
$\Lambda_{P_+ +1} = Q_{RNS}(c\xi\p\xi e^{-2\phi}\Lambda_{P_+ +1})$
and $\Lambda_{P_+} = \eta_0(\xi\Lambda_{P_+})$.
Plugging into 
\whg, one finds
$$U= Q_{RNS}\eta_0 (  
-c\xi\p\xi e^{-2\phi}\Lambda_{P_+ +1} + \xi \Lambda_{P_+}),$$
which implies
that $U$ would have been pure gauge using the definition of \phys.

So any state in the cohomology of $Q' = Q_{RNS} + \eta_0$ determines
a physical vertex operator using the definition of \phys. Also,
any two physical vertex operators which are related by the picture
changing operations of \raising\ or \lowering\ are described by
the same state in the cohomology of $Q'$. For example, the vertex
operator $U$ and the picture-raised operator $\{Q_{RNS},\xi U\}$
are related by the gauge transformation $\d U = Q'(\xi U)$.
Similarly, the vertex operator $U$ and the picture-lowered operator
$\{\eta_0,c\xi\p\xi e^{-2\phi} U\}$ are related by the gauge
transformation $\d U = Q' (c\xi\p\xi e^{-2\phi} U)$. 

\newsec{Relating the RNS and Pure Spinor BRST Operators}

In this section, the operator $Q'=Q_{RNS}+\oint \eta$ of section 3
will be related to the operator $Q=\oint \l^\a d_\a$ of section 2.
This will be done by first finding a field redefinition which maps
$Q'$ to a spacetime supersymmetric operator $Q_{U(5)}$, and then
constructing a similarity transformation $R$ which satisfies
$e^{-R} Q_{U(5)} e^R= Q$.

Because $\{q^{-\half}_\a, q^{-\half}_\b\} = \g^m_{\a\b}
\oint e^{-\phi}\psi_m$ in the RNS formalism where $q_\a^{-\half}=
\oint e^{-\half\phi} \Sigma_\a$ is the spacetime supersymmetry generator
in the $-\half$ picture and $\Sigma_\a$ is the spin field, 
the spacetime supersymmetry algebra only
closes up to picture changing in the RNS formalism. Nevertheless, 
since
$\{q^{-\half}_\a, q^{+\half}_\b\} = \g^m_{\a\b} \oint \p x_m$ where
$q^{+\half}_\b = \oint (b \eta e^{{3\over 2}\phi} \Sigma_\b +
e^{\half\phi} \g^m_{\a\b} \Sigma^\a \p x_m)$ is the
spacetime supersymmetry generator in the $+\half$ picture, 
one can make a subset of the algebra close 
by choosing some of the supersymmetry generators in the $-\half$ picture
and others in the $+\half$ picture. However, this choice
necessarily breaks manifest Lorentz
invariance. As shown in \ufive\ and will be
reviewed in the following subsection, $U(5)$ is the maximum
subgroup of the (Wick-rotated) $SO(10)$ Lorentz group
which can be manifestly preserved, which is the subgroup of
$SO(10)$ that leaves a pure spinor invariant.
Other choices for the subgroup are
useful for describing Calabi-Yau compactifications of the superstring
\ref\four{N. Berkovits, {\it Covariant Quantization of the 
Green-Schwarz
Superstring in a Calabi-Yau Background}, Nucl. Phys. B431 (1994) 258,
hep-th/9404162.}\top.

\subsec{Review of $U(5)$ Formalism}

Under $U(5)$, the $SO(10)$ spinor $q_\a$ splits into 
$(q_+, q^{ab}, q_a)$ which transform as $(1_{-{5\over 2}}, 10_{-\half},
\overline 5_{3\over 2})$ representations of $SU(5)_{U(1)}$.
If $q_+$ is chosen in the $+\half$ picture
and $q_a$ is chosen in the $-\half$ picture, these six generators preserve
the
supersymmetry algebra since $\{q_a,q_b\} = \{q_+,q_+\}=0.$

With this choice of picture, it is natural to define $\t^a=e^{\half\phi}
\Sigma^a$ and $\t^+ = c\xi e^{-{3\over 2}\phi} \Sigma^+$ so that
$\{q_a,\t^b\}=\d_a^b$ and $\{q_+,\t^+\}=1$.
One then defines conjugate momenta to $\t^a$ and $\t^+$ by
$p_a = e^{-\half\phi}\Sigma_a$ and $p_+ = b\eta e^{{3\over 2}\phi}\Sigma_+$.
Since the RNS variables include twelve fermions $(\psi^m,b,c)$ and
two chiral bosons $(\beta,\gamma)$, there are still two independent
chiral bosons which will be defined as\foot{These two chiral bosons
are related to $\rho$ and $\sigma$ of \ufive\ by $s=\half(-\rho+i\sigma)$
and $t= \half(\rho+i\sigma)$.}
$$\p s = -bc -{3\over 2}\p\phi -\half\sum_{a=1}^5\psi^{2a-2}\psi^{2a-1},\quad 
\p t = -\xi\eta
 +{3\over 2}\p\phi +\half\sum_{a=1}^5\psi^{2a-2}\psi^{2a-1}.$$

{}From the RNS OPE's, one finds that the only singular OPE's of
$(x^m,\t^+,\t^a,p_+,p_a,s,t)$ are
$$x^m(y) x^n(z) \to -\eta^{mn} \log|y-z|^2,$$
$$\t^+(y) p_+(z) \to (y-z)^{-1},\quad
\t^a(y) p_b(z) \to \d^a_b (y-z)^{-1},\quad
s(y) t(z) \to \log (y-z).$$
In terms of these GS-like variables, the RNS stress tensor is
\eqn\stress{T_{RNS} = -
\half \p x^m \p x_m - p_+\p \t^+ - p_a\p \t^a + \p s\p t + \p^2 s,}
the RNS BRST operator is
\eqn\rnsbrst{Q_{RNS}= \oint 
( e^t [ p_a \p x^a - \t^+ \p x_a \p x^a
+ p_a \t^+ \p\t^a
+ \p\t^+ (\p s + \half\p t) ]  }
$$- {1\over{120}}e^{2t-s}  \e^{abcde} ( p_a p_b
p_c p_d p_e - 5 \t^+ p_a p_b p_c p_d \p x_e ) ),$$
the RNS $b$ and $\eta$ ghosts are 
\eqn\bghost{b_{RNS}= e^{-t} p_+,\quad
\eta = e^{s} p_+,}
and the RNS ghost-number current is 
\eqn\rnsj{j_{RNS} = cb +\eta\xi = \p s + \p t, }
where 
the $SO(10)$ vector $x^m$ has been split into its $U(5)$ components
as $x^a$ and $x_a$ which transform as $5_{+1}$ and $\overline 5_{-1}$
representations of $SU(5)_{U(1)}$ and satisfy the OPE
$x^a(y) x_b(z)\to -\d_a^b \log|y-z|^2$.

To relate $Q' = Q_{RNS} +\oint\eta$ with $Q=\oint \l^\a d_\a$, it will
be convenient to first perform a unitary transformation
on the
GS-like variables $(p_+,p_a,x^a)$ such that
\eqn\unit{ p_+^{new}=  p_+^{old} +\half \t^a \p x_a, 
\quad p_a^{new}= p_a^{old} -\half \t^+ \p x_a,
\quad x^a_{new} = x^a_{old} - \half\t^a \t^+.} 
In terms of the ``new''
GS-like variables,
one can check that 
$$Q' = Q_{RNS}+\oint\eta $$
\eqn\Qu{
= \oint (e^{s}\widehat d_+ + 
 e^t [ \widehat d_a \widehat \Pi^a 
+ \p\t^+ (\p s -{3\over 4}\p t) ]
 - {1\over{120}}e^{2t-s}  \e^{abcde} \widehat d_a\widehat d_b
\widehat d_c\widehat d_d\widehat d_e) }
where 
\eqn\hatted{\widehat d_+= p_+ - \half \t^a \p x_a, \quad
\widehat d_a = p_a -\half \t^+ \p x_a,\quad \widehat\Pi^a = 
\p x^a +\half( \t^a \p\t^+ + \t^+ \p\t^a), }
are defined like $d_\a$ and $\Pi_m$ in \ddef\ and \oped\ but with
$\t_{ab}$ set to zero. The operator of \Qu\ is manifestly
invariant
under the six supersymmetry transformations
generated by $\widehat q_a=\oint (p_a +\half \t^+ \p x_a)$ and 
$\widehat q_+ = \oint(p_+ + \half \t^a \p x_a)$.

So using the arguments of section 3, physical RNS states can be
described in $U(5)$ language as states in the cohomology of the
operator \Qu.
After adding a topological sector to the $U(5)$ formalism, 
it will be shown in the next subsection
how to relate $Q'$ with the pure spinor BRST
operator $Q=\oint\l^\a d_\a$.

\subsec{Supersymmetric $U(5)$ formalism}

The cohomology of $Q'$ of \Qu\ 
defines physical states in a manner which manifestly preserves
six spacetime supersymmetries.
As will now be shown,
all sixteen supersymmetries can be made manifest if one
adds ten new fermionic variables and their
conjugates, $(\t_{ab},p^{ab})$, as well as ten new bosonic variables
and their conjugates, $(u_{ab},v^{ab})$, to the $U(5)$
variables of the previous subsection. These new variables are
not related to RNS worldsheet variables so the
BRST operator must be modified such that the new variables do not
affect the physical states. Note that a similar trick was used
in the six-dimensional version of the hybrid formalism where four
$\t$ variables and their conjugates were added in order to make
all eight spacetime supersymmetries manifest \ref\sixcur{N. Berkovits,
{\it Quantization of the Type II Superstring in
a Curved Six-Dimensional Background}, Nucl. Phys. B565 (2000) 333,
hep-th/9908041.}.

The first step is to change the RNS action to
$$ S_{U(5)} = S_{RNS} + \half
\int d^2 z (  p^{ab}\overline\p\t_{ab} - v^{ab}\overline\p u_{ab} )$$
so that the new variables satisfy the OPE's
$$p^{ab} (y) \t_{cd}(z) \to \d^{[a}_c \d^{b]}_d (y-z)^{-1},\quad
v^{ab} (y) u_{cd}(z) \to \d^{[a}_c \d^{b]}_d (y-z)^{-1}.$$
One now modifies the BRST operator to
\eqn\quf{Q_{U(5)} = \l^\a d_\a + e^t(d_a\Pi^a + \p\t^+ (\p s-{3\over 4}\p t)
) + {1\over {12}}e^{t-s}\e^{abcde}u_{ab} d_c d_d d_e - e^{2t-s} (d)^5 }
where $\l^\a$ is defined in \paramone,
$d_\a$ and $\Pi^m$ are defined in \ddef\ and \oped, 
and $(d)^5 = {1\over{120}}
\e^{abcde} d_a d_b d_c d_d d_e$. Using the OPE's
of \oped, one can check that $Q_{U(5)}$ of \quf\ is nilpotent.\foot
{A useful trick for computations is to perform a unitary transformation
analogous to \unit\ such that $d_a=p_a$. This unitary transformation 
changes only the term $e^t\p\t^+ (\p s-{3\over 4}\p t)$ to
$e^t\p\t^+ (\p s-2\p t)$ in $Q_{U(5)}$ and simplifies OPE's involving
$d_a$. For example, after the transformation,
$d_+(y) d_a d_b(z) \to (y-z)^{-2}\p\t_{ab}(y)-(y-z)^{-1}\Pi_{[a}(y)
d_{b]}(z) + d_+ d_a d_b (z).$}
It will also be convenient to modify the $b$ ghost and ghost-number current
of \bghost\ and \rnsj\ to
\eqn\bpure{b_{U(5)} = e^{-t} d_+ + \half v^{ab}\p\t_{ab}, \quad
j_{U(5)} = \p s+\p t + \half u_{ab} v^{ab},}
so the
stress tensor of \stress\ is modified to
$$T_{U(5)} = \{Q_{U(5)}, b_{U(5)}\}
= T_{RNS} - \half p^{ab}\p\t_{ab} + \half v^{ab}\p u_{ab} $$
\eqn\stressuf{ = -\half \p x^m \p x_m - p_\a \p\t^\a + \half v^{ab}\p u_{ab}
+ \p t\p s + \p^2 s.}

It will now be shown that
$Q_{U(5)}$ in the enlarged Hilbert space
has the same cohomology as $Q'$ of \Qu\ in the old
Hilbert space without the new variables.
To relate the $Q_{U(5)}$ and $Q'$ cohomologies, first write
\eqn\firstw{Q_{U(5)} = \half\oint u_{ab}p^{ab} + Q'  + 
f(u_{ab},\t_{ab})}
where $f(u_{ab},\t_{ab})$ includes all terms in $Q_{U(5)}$ except
$\half\oint u_{ab}p^{ab}$ which involve $u_{ab}$ or $\t_{ab}$.
If $(p^{ab},\t_{ab})$ are assigned ``charge''
$(-2,+2)$, $(v^{ab},u_{ab})$ are assigned ``charge'' $(-1,+1)$, and
all other variables are assigned zero ``charge'', then the terms
of \firstw\ are written in order of increasing ``charge'', i.e.
$Q_{U(5)} = Q_{(-1)} + Q_{(0)} + 
Q_{(1)}+ ...$
where $Q_{(n)}$ carries ``charge'' $n$.\foot{I would like to
thank Edward Witten for suggesting this method for analyzing the cohomology.} 

Now consider a vertex operator 
$U_{U(5)}= \sum_{n=C}^\infty U_{(n)}$
where $n$ denotes the ``charge'' of the term $U_{(n)}$ and $U_{(C)}$
is the term of lowest ``charge''. Note that $C$ is bounded from below
since only variables of positive conformal weight
carry negative ``charge''. Suppose that $U_{U(5)}$ is
in the cohomology of $Q_{U(5)}$, i.e. $Q_{U(5)} U_{U(5)}=0$ and
$\d U_{U(5)}= Q_{U(5)}\Lambda$ for $\Lambda= \sum_n \Lambda_{(n)}$. 
Then $Q_{(-1)} U_{(C)}=0$ and
$\d U_{(C)} = Q_{(-1)} \Lambda_{(C+1)}$ where $Q_{(-1)}=
\half\oint u_{ab} p^{ab}$. So using the quartet mechanism, $U_{(C)}$
can be gauge-fixed to be independent of the new variables $(u_{ab},v^{ab},
\t_{ab},p^{ab})$. But since ``charge'' is carried only by 
the new variables, one can choose a
gauge such that $U_{(C)}=0$ if $C<0$. Similarly, one can choose a gauge
such that $U_{(n)}=0$ for all $n<0$ and such that $U_{(0)}$ is independent
of the new variables.

In this gauge, $Q_{(0)} U_{(0)}=0$ and
$\d U_{(0)}=Q_{(0)} \Lambda_{(0)}$ where $U_{(0)}$ and
$\Lambda_{(0)}$ are independent of the new variables and $Q_{(0)}=Q'$. 
In other words,
$U_{(0)}$ describes states in the cohomology of $Q'$ in the Hilbert
space without the new variables. Finally, it will be shown by induction
that all terms $U_{(n)}$ for $n>0$ are determined by $U_{(0)}$ up
to a gauge transformation.

Suppose that $U_{(n)}$ is known for $0\leq n\leq M$. Then 
$Q_{(-1)}Q_{U(5)} U_{U(5)}=0$ implies that 
$Q_{(-1)} (\sum_{n=0}^M Q_{(n)} U_{(M-n)}) =0.$
But since $Q_{(-1)} = \half\oint u_{ab} p^{ab}$ has trivial
cohomology at non-zero ``charge'', there exists an operator
$U_{(M+1)}$ satisfying
$Q_{(-1)} U_{(M+1)} = - \sum_{n=0}^M Q_{(n)} U_{(M-n)}$
which is uniquely determined up to the gauge transformation
$\d U_{(M+1)}= Q_{(-1)} \Lambda_{(M+2)}$. Similarly, all
terms $U_{(n)}$ for $n>0$ are determined by $U_{(0)}$ up
to a gauge transformation. This completes the proof that the cohomology
of $Q_{U(5)}$ of \quf\ in the enlarged Hilbert space is equivalent to the
cohomology of $Q'$ of \Qu\ in the Hilbert space without the new variables.

\subsec{Similarity transformation}

The next step in relating $Q'$ and $Q=\oint \l^\a d_\a$
is to find a similarity transformation $R$
such that $e^{-R} Q_{U(5)} e^{R}=\oint\l^\a d_\a$ where $\l^\a$ is
defined in \paramone.
This similarity transformation is
\eqn\simrone{R=\oint (e^{t+s} g_a \Pi^a -{1\over 4}
\e^{abcde} e^t g_a u_{bc} d_d d_e)}
where $g_a$ is defined to be any function of $u_{ab}$ which 
satisfies 
\eqn\defea{{1\over 8}e^{abcde} g_a u_{bc} u_{de} =1.}
To preserve \defea, $g_a$ is defined to satisfy the OPE
\eqn\gope{g_a(y) v^{bc}(z) \to \half(y-z)^{-1} \e^{bcdef} g_a g_d u_{ef}(z).}
For example, if
$\e^{abcd5} u_{ab} u_{cd}$
is non-zero, one can choose $g_a = 8~\d_a^5 (\e^{bcde5} u_{bc}u_{de})^{-1}$, 
which can be checked to satisfy \gope.
Note that \defea\
has no solutions if
all five components of $\e^{abcde}u_{bc} u_{de}$ are zero, i.e.
if $\l^a =0$. As will
be discussed in subsection (5.2), this creates subtleties when
using the similarity transformation of $R$ to relate $U(5)$ and
pure spinor vertex operators.\foot{These subtleties are reminiscent
of the subtleties with the $\l^\a$ parameterization of \paramone\
when $\l^+ =0$.
Perhaps these subtleties can be avoided by using transition functions
to relate different patches of $\l^\a$ space in a manner analogous
to the conventional treatment of Penrose's twistor space \ref\sc{S. 
Cherkis, private communication.}.}
To check that $e^{-R} Q_{U(5)} e^R = \oint \l^\a d_\a$, note that
\eqn\commr{[R,\l^\a d_\a] = e^t(d_a\Pi^a+\p\t^+(\p s-{3\over 4}
\p t))+{1\over{12}}e^{t-s}\e^{abcde}u_{ab} d_c d_d d_e,}
$$[R,[R,\l^\a d_\a]~] =-2 e^{2t-s} (d)^5,\quad 
{\rm and} ~~~[R,[R,[R,\l^\a d_\a]~]~]=0.$$

So the RNS operator $Q' = Q_{RNS} +\oint\eta$ has been mapped to the
pure spinor operator $Q=\oint\l^\a d_\a$ using a field redefinition
and similarity transformation. In the next section, this map
will be used to relate vertex operators and tree amplitudes in the 
two formalisms. 

\newsec{Relating the RNS and Pure Spinor Vertex Operators}

\subsec{Relating RNS and $U(5)$ vertex operators}

As discussed in subsections (4.1) and (4.2), the RNS worldsheet variables
can be related by a field redefinition to 
the GS-like
variables $(x^m,\t^+,\t^a,p_+,p_a,s,t)$, which 
can then be covariantized
by adding the ``topological''
variables $(\t_{ab},p^{ab},u_{ab},v^{ab})$. 
Physical states in this $U(5)$ Hilbert space are defined
as states in the cohomology of $Q_{U(5)}$ of \quf. 

As was shown in subsection (4.2), 
any physical RNS vertex operator $U_{RNS}$ can be mapped
to a vertex operator $U_{U(5)}$ in the cohomology of $Q_{U(5)}$ by 
using the field redefinition
to write $U_{RNS}$ in terms of $U(5)$ variables and defining
$U_{U(5)} =  U_{RNS} + ...$ where $...$ involves operators of
positive ``charge'' which are determined by $U_{RNS}$ up to
a gauge transformation. 
Similarly, any physical vertex operator $U_{U(5)}$ can be mapped
to a vertex operator $U_{RNS}$ in the cohomology of $Q'=Q_{RNS}
+\oint \eta$ by first
choosing a gauge in which $U_{U(5)}$ has no operators with negative
``charge''. One can then define
$U_{RNS}$ as the
operator in $U_{U(5)}$ with zero ``charge'' and use
the field redefinition
to write this operator in terms of RNS variables.

In order to complete the map between physical RNS and physical pure
spinor vertex operators, one therefore needs to find a map between
physical
$U(5)$ vertex operators and pure spinor vertex operators in
the cohomology of $Q=\oint \l^\a d_\a$. 

\subsec{Relating $U(5)$ and pure spinor vertex operators}

Although $Q= e^{-R}Q_{U(5)} e^{R}$ where $R$
is defined in \simrone, this similarity transformation
cannot be directly used to relate physical pure spinor and $U(5)$ vertex
operators by $U=e^{-R} U_{U(5)} e^{R}$ and $U_{U(5)}=e^R U e^{-R}$
since $R$ does not preserve the relevant Hilbert spaces.
Note that both the $U(5)$ and pure spinor Hilbert spaces 
consist of
functions constructed from the variables $(x^m,\t^\a,p_\a,s,t,u_{ab},
v^{ab})$. However, 
vertex operators in the pure spinor
Hilbert space are required to be Lorentz-covariant functions of $(s,t,u_{ab},
v^{ab})$, i.e. functions which are polynomials in $\l^\a, N_{mn}$ 
and $\p h$ of \paramone, \lorentz, and \hdef.
But $e^{-R} U e^R$ does not necessarily have this property since 
$Q_{U(5)}$ and $R$ are not Lorentz invariant. Also,
vertex operators in the $U(5)$ Hilbert space must be independent of
$g_a$ so that they are well-defined when $\e^{abcde} u_{bc} u_{de}=0$.
So because of explicit $g_a$ dependence in $R$, $e^R U e^{-R}$ is
not necessarily an allowable $U(5)$ vertex operator.

Although $Q_{U(5)}$ and $R$ are not super-Poincar\'e invariant,
they are manifestly invariant under 
all sixteen spacetime supersymmetry transformations and under a
$U(5)$ subgroup of the (Wick-rotated) $SO(10)$ Lorentz transformations.
It
will be convenient to define operators which transform
covariantly under this subgroup of super-Poincar\'e transformations as
``almost super-Poincar\'e covariant'' operators, or ASPC operators
for short. 

It will now be conjectured that for each physical pure spinor vertex
operator $U$, one can define a physical $U(5)$ vertex operator
$U_{U(5)}$ by
\eqn\conjone{U_{U(5)} = e^{R} U e^{-R} + Q_{U(5)} \Omega}
where $\Omega$ is some ASPC operator which is allowed to depend on $g_a$.
Furthermore, it will be 
conjectured that different ASPC choices for $\Omega$ only
change $U_{U(5)}$ by a $U(5)$ gauge transformation, i.e. by
$Q_{U(5)} \Lambda_{U(5)}$ where $\Lambda_{U(5)}$ is independent of $g_a$.

It is crucial that $\Omega$ is restricted to be an ASPC
operator since otherwise, the map could relate physical vertex operators
with BRST-trivial vertex operators. For example, $Q_{U(5)} U_{U(5)}=0$ implies
that $U_{U(5)} = Q_{U(5)} (-e^s\t^a g_a U_{U(5)})$
and $Q U=0$ implies that $U= Q (e^{-s}\t^+ U)$.
But 
$e^s\t^a g_a U_{U(5)}$ and
$e^{-s} \t^+ U$ are not ASPC operators
since they contain explicit dependence on $\t^\a$ zero modes.
Note that to transform covariantly under spacetime supersymmetry,
the ASPC operator $\Omega$ must be constructed from products
of spacetime-supersymmetric operators and covariantly
transforming spacetime superfields
which appear in $U$.

Since $R$ is proportional to $e^t$ and and pure spinor
operators $U$ have 
no $e^t$ dependence, one can always choose
$\Omega$ of \conjone\ such that it has only positive powers of $e^t$.
In this case, 
\eqn\specia{U_{U(5)} = U + U_{ASPC}}
where $U$ is a physical pure spinor vertex operator and
$U_{ASPC}$ is an ASPC operator which is independent of $g_a$
and contains only positive powers of $e^t$. 
Any physical $U(5)$ vertex operator
of the form of \specia\ will be called a ``special'' $U(5)$
vertex operator. 

Since the RNS and pure spinor cohomologies were
shown to be equivalent in \cohom,
the conjecture of \conjone\ implies that any state in the cohomology
of $Q_{U(5)}$ can be represented by a ``special'' $U(5)$ vertex 
operator.\foot{More precisely, the
cohomologies at non-zero $P_+$ were shown to be equivalent. As will
be discussed in section 6, the RNS cohomology has an extra zero-momentum
state at ghost-number $-1$ which cannot be represented by a
``special'' $U(5)$ vertex operator.} 
Furthermore, the conjecture implies that any two ``special'' vertex operators
with the same pure spinor vertex operator as their $e^t$-independent
component are related by a $U(5)$
gauge transformation. So the conjecture of \conjone\
not only implies a map from
physical pure spinor vertex operators
to physical $U(5)$ vertex operators, but also implies
(with the exception of the zero momentum
state mentioned in footnote 13) a map using \specia\ from
physical
$U(5)$ vertex operators to physical pure spinor vertex operators. 
 
Evidence for the conjecture of \conjone\ will now be obtained
by explicitly constructing ``special'' $U(5)$ vertex operators
for the physical massless states
and by using the property that  
``special'' massive vertex operators can be obtained from the OPE's
of ``special'' massless vertex operators. 

\subsec{``Special'' massless vertex operators}

In the pure spinor formalism where $Q=\oint\l^\a d_\a$, 
the physical unintegrated massless
vertex operator of ghost-number $+1$ is $U=\l^\a A_\a(x,\t)$ where
$D_{(\a} A_{\b)} = \g^m_{\a\b} B_m$ and $D_\a B^m-\p^m A_\a
= \g_{\a\b}^m W^\b$.
To compute $U_{U(5)}$, it will be convenient to first use the gauge
invariance $\d A_\a = D_\a \Lambda$ to gauge-fix
$A_a=0$. In this gauge,
\eqn\spem{e^R (\l^\a A_\a) e^{-R} = e^R (e^{s} A_+ +\half
u_{ab} A^{ab}) e^{-R}}
$$= e^s A_+ +\half
u_{ab} A^{ab} -\half \e^{abcde} e^t g_a u_{bc} d_d u_{ef} B^f
-{1\over 4}\e^{abcde} \p (e^t g_a u_{bc}) u_{de} W^+ $$
$$+ e^{s+t} ({1\over 4}
\e^{abcde} g_a u_{bc} W_{de} +\half g_a u_{bc} \p^a A^{bc})
+ {1\over 4}e^{2t} \e^{abcde} g_a u_{bc} d_d d_e W^+ + 
e^{2t+s} g_a \p^a W^+$$
$$= e^s A_+ + \half
u_{ab} A^{ab} + e^t (d_a B^a + \p(s- 2t)W^+)
 + Q_{U(5)}(e^{t+s} g_a B^a),$$ 
where $A_\a$, $B_m$ and $W^\a$ have been written in terms of their
$U(5)$-irreducible components.

So for massless states, the conjecture of \conjone\
has been confirmed where
in the gauge $A_a=0$, 
$U_{U(5)} = \l^\a A_\a
 + e^t (d_a B^a + \p(s-2t)W^+)$ and $\Omega = e^{t+s}g_a B^a$.
To construct $U_{U(5)}$ in other gauges, use the fact that 
$\d U_{U(5)} = Q_{U(5)}\Lambda$ where $\d A_\a = D_\a \Lambda$ and
$\d B_m = \p_m \Lambda$ to learn that 
\eqn\vcov{U_{U(5)} = \l^\a A_\a
 + e^t (d_a B^a +  \p(s-2t)W^+)
+ \lim_{y\to z}  :e^t \Pi^a(y) A_a(z):  } 
$$+\lim_{y\to z}:{1\over 4}e^{t-s} \e^{abcde} u_{ab} d_c d_d (y)A_e (z):
-\lim_{y\to z}
:{1\over 24}e^{2t-s} \e^{abcde} d_a d_b d_c d_d(y) A_e(z): $$
where $\lim_{y\to z}:f(y) g(z): = {1\over{2\pi i}}
\oint dy (y-z)^{-1} f(y) g(z)$.
Note that up to normal ordering, $U_{U(5)}$ of \vcov\
can be obtained from $Q_{U(5)}$ of \quf\ by replacing 
\eqn\replc{d_\a \to d_\a + A_\a, \quad \Pi_m \to \Pi_m + B_m,~~{\rm and}~~~
\p\t^\a \to \p\t^\a + W^\a,}
as was suggested in \siegel.

As will be shown in subsection (6.2), the massless integrated $U(5)$ vertex
operator $V_{U(5)}$ can be obtained from $U_{U(5)}$ of \vcov\ by
\eqn\defintv{\int V_{U(5)} = \int [(b_{U(5)})_{-1} U_{U(5)} +
Q_{U(5)}( (b_{U(5)})_0 V)]}
where $(b_{U(5)})_{-1}$ and
$(b_{U(5)})_0$ are modes of
the $U(5)$ $b$ ghost of $(6.1)$ and $V$ is the pure
spinor integrated vertex operator of \supermax. 
One can check that $V_{U(5)}$
is a ``special'' $U(5)$
vertex operator satisfying $Q_{U(5)} V_{U(5)}= \p U_{U(5)}$
whose $e^t$-independent component is $V$ of \supermax.

By taking suitable OPE's of $V_{U(5)}$ and $U_{U(5)}$ of \defintv\ and 
\vcov,
one can also construct ``special'' $U(5)$
vertex operators for massive states.
For example, vertex operators with $(mass)^2=N$ can be obtained by
taking the contour integral of $V_{U(5)}$ around $U_{U(5)}$ where
the momenta $k^m$ and $l^m$ of these two vertex operators are 
chosen to satisfy $(k+l)^2 =  N$.
It might be possible to use such a construction to confirm the
conjecture of \conjone\ for all massive states.

\subsec{Equivalence of tree amplitudes}

Using the results of the previous subsections, it will now
be shown that the RNS and pure spinor tree amplitude prescriptions
are equivalent. Using RNS vertex operators in the cohomology of
$Q'=Q_{RNS}+\oint\eta$, $N$-point tree amplitudes can be defined by computing
the correlation function 
\eqn\amplrns{A_{RNS} = \int dz_4 ... dz_N \langle ~U_1(z_1) ~ U_2(z_2) ~
U_3(z_3) ~\int dz_4 V_4(z_4) ... \int dz_N V_N(z_N) ~\rangle}
where $U_r$ are dimension-zero vertex operators satisfying
$Q' U_r=0$ and $V_r$ are dimension-one vertex operators satisfying
$Q' V_r = \p U_r$. 

Normally, one uses the normalization that $\langle c \p c \p^2 c e^{-2\phi}
\rangle =1$, but such a normalization would break gauge invariance since
it implies that the amplitude vanishes unless the sum of the pictures
of the vertex operators is $-2$. So it will instead be convenient to
use the normalization prescription that 
\eqn\normnew{\langle c \p c \p^2 c e^{-2\phi} + Q'\Lambda \rangle =1}
for
any gauge parameter $\Lambda$. For example, 
$\langle c \p c \eta \rangle =$
$\langle c\p c \p^2 c e^{-2\phi}
+ Q' (\xi c \p c\p^2 c e^{-2\phi}) \rangle =1$ using
this normalization.
Note that this normalization
prescription is invariant under picture-changing
and agrees with the standard prescription when the sum of the pictures
of the vertex operators is $-2$.

To explicitly compute \amplrns\ with the normalization of \normnew,
use the free field OPE's to write $A_{RNS} = \int
dz_4 ... dz_N \langle f(z_1) \rangle$ where $f(z_1)$ is some operator
located at the point $z_1$. Since the external vertex operators are
BRST invariant and $c\p c\p^2 c e^{-2\phi}$ is the only non-trivial
element in the cohomology of $Q'$ at ghost-number $+3$,
$f= c \p c \p^2 c e^{-2\phi} F(z_r,k_r, \eta_r) + Q' (\Lambda(z_1))$
for some gauge parameter $\Lambda(z_1)$ and for
some function $F(z_r,k_r,\eta_r)$
which depends on the external momenta and polarizations.
So $A_{RNS} = \int dz_4 ... dz_N F(z_r,k_r,\eta_r)$.

To compare with the pure spinor tree amplitude prescription of
subsection (2.3), use the
results of the previous subsections to map the RNS vertex operators
of \amplrns\ 
into the pure spinor vertex operators of \ampl. 
Under this map, one can check
that the RNS operator $c\p c \p^2 c e^{-2\phi}$ gets mapped into
the pure spinor operator $(\t\g^{mnp}\t)(\l\g_m\t)(\l\g_n\t)(\l\g_p\t)$,
which is the unique ghost-number $+3$ element in the cohomology of 
$Q=\oint \l^\a d_\a$. So the RNS normalization of \normnew\ coincides
with the normalization prescription of \zeromode, 
implying that the tree amplitude prescriptions are equivalent.

\newsec{Pure Spinor $b$ Ghost}

As mentioned earlier, there is a ghost-number $-1$ operator in the
RNS cohomology, $\int d^2 z ~\mu(z)~ b_{RNS}(z)$ where $\mu(z)$
is a Beltrami differential, which has no super-Poincar\'e
invariant counterpart in the pure spinor cohomology.
This operator is in the RNS cohomology because its
anticommutator with $Q_{RNS}$ is $\int d^2 z ~\mu(z) ~T_{RNS}(z)$, which
is a total derivative in the moduli space of Riemann surfaces.

Although there are no super-Poincar\'e covariant operators of negative
ghost number, the conjecture of \conjone\
implies that the OPE of this operator
with any physical state of positive ghost number can be expressed
as a super-Poincar\'e covariant pure spinor vertex operator. 
As will now be shown, one can define a pure spinor version of the
$b$ ghost which, although not super-Poincar\'e invariant, can
be used to construct pure spinor integrated vertex operators from
unintegrated operators, to convert pure spinor string
antifields into fields,
and to define pure spinor tree amplitudes in a worldsheet reparameterization
invariant manner. It is hoped that this operator will also be useful
for defining loop amplitudes using the pure spinor formalism. 

\subsec{Construction of the pure spinor $b$ ghost}

To express the $b$ ghost in pure spinor language, it is useful
to first remove
all negative powers of $e^t$
from the $U(5)$ version of the $b_{RNS}$ ghost of \bpure\ by
adding the BRST-trivial operator 
$Q_{U(5)} (-e^{-s-t}).$ With this modification,
\eqn\buf{b_{U(5)} = 
 e^{-t} d_+ + \half v^{ab} \p\t_{ab}
+ Q_{U(5)}(-e^{-s-t})} 
$$ =  e^{-s} (d_a \Pi^a + \p(s-t)\p\t^+ -{1\over 4}\p^2\t^+) 
+ \half v^{ab}\p\t_{ab}
+ e^{t-2s} (d)^5. $$

Although $b_{U(5)}$ is an ASPC operator, its $e^t$-independent
component is not Lorentz invariant so $b_{U(5)}$ is not
a ``special'' $U(5)$ vertex operator as defined in \specia. 
Nevertheless, it will be useful
to define the pure spinor version of the $b$ ghost to be the
$e^t$-independent component of \buf, i.e.
\eqn\bdef{b 
=  e^{-s} (d_a \Pi^a + \p(s-t)\p\t^+ -{1\over 4}\p^2\t^+) 
+ \half v^{ab}\p\t_{ab}.}
It is interesting to 
note that $b$ can also be written as 
$\lim_{y\to z}:e^{-s(y)} b^+(z):$ where $b^+$
is the $1_{{5\over 2}}$ component of the covariantly transforming
spinor
\eqn\bspinor{b^\alpha =\half \g^{\a\b}_m d_\b \Pi^m + {1\over 4}
(\g^{mn})^\a{}_\b
N_{mn} \p\t^\b  -{1\over 4} \p^2 \t^\a +\half \p h \p\t^\a}
and $\lim_{y\to z} :~:$ is defined as in \vcov.
One can check that $\{Q,b^\a\}= \lim_{y\to z}:\l^\a(y) T(z):$ 
where $T$ is defined in \stress, so $\{Q,b\}=T$ as desired. 

If $U_{U(5)}$ is a physical $U(5)$ vertex operator of positive ghost
number $N$, then $(\int \mu~ b_{U(5)}) U_{U(5)}$ is an
ASPC physical vertex operator of ghost number $N-1$. Although 
$(\int \mu~ b_{U(5)}) U_{U(5)}$ is not ``special'' since its
$e^{t}$-independent component is not a pure spinor vertex operator, 
the conjecture of \conjone\
implies that it is related to a ``special'' $U(5)$
vertex operator $V_{U(5)}$ by an ASPC gauge transformation, i.e. 
\eqn\impl{V_{U(5)} = 
(\int \mu~b_{U(5)}) U_{U(5)} + Q_{U(5)}\Lambda_{U(5)}}
where $\Lambda_{U(5)}$ is an ASPC operator. Taking the $e^t$-independent
component of \impl, one learns that for any pure spinor vertex operator
$U$ at ghost number $N$, there exists an ASPC operator $\Lambda$ such that
\eqn\pureimp{V = (\int \mu~ b ) U + Q\Lambda}
where $V$ is a physical pure spinor vertex operator of 
ghost number $N-1$.\foot{
Note that the RNS and pure spinor ghost numbers coincide since
after including the topological sector, the RNS ghost charge
is $\oint (cb+\eta\xi+\half u_{ab} v^{ab}) =
\oint (\p s + \p t + \half u_{ab} v^{ab})$. When acting on pure
spinor states (which have no $e^t$ dependence), this ghost charge
is equivalent to the pure spinor ghost charge
$2\oint \p h = \oint (-3\p s + \p t + \half u_{ab} v^{ab})$ which
was defined in subsection (2.1).}
This property will now be used to relate physical pure spinor
vertex operators of different ghost numbers.

\subsec{Integrated vertex operators from unintegrated operators}

In bosonic and RNS string theory, one can use the $b_{-1}$ mode of the 
$b$ ghost to construct ghost-number zero integrated vertex operators
from ghost-number $+1$ unintegrated operators. The only new feature
in the pure spinor formalism is that $b$ is not super-Poincar\'e
invariant, so one needs to perform an ASPC gauge
transformation as in \pureimp\ in order that the resulting
integrated operator is super-Poincar\'e covariant. 

It will now be shown that when there exists a super-Poincar\'e covariant
operator $V$ which is a dimension one primary field\foot
{For the dimension one massless vertex operator of \supermax, $V$ is primary
when $A_\a$ is in Lorentz gauge, i.e.
$\g_m^{\a\b}\p^m D_\a A_\b =0$.}
satisfying $QV=\p U$,
the ASPC gauge parameter $\Lambda$ satisfying \pureimp\
can
be chosen as $\Lambda =\int b_0 V$. To prove this,
note that 
\eqn\pron{\int Q(b_0 V) = \int (T_0 V + b_0 \p U) = \int (V - b_{-1} U +
\p (b_0 U)) =\int (V - b_{-1} U)}
where $T_0$ is the zero mode of the stress tensor which satisfies
$T_0 V=V$ for dimension one primary fields.
So $\int V =\int b_{-1} U + Q\Lambda$ as desired.

\subsec{Fields and antifields}

In bosonic and RNS string theory, the presence of the 
$b$ zero mode implies that there is a doubling of the cohomology
at ghost numbers $+1$ and $+2$ where the ghost number $+1$ states
are associated with the string field $U$ and the ghost number $+2$ states
are associated with the string antifield $U^*$. In the gauge where the
antifield $U^*$ is annihilated by $T_0$, $b_0 U^*$ is a physical
ghost number $+1$ field. This procedure can be applied to the pure
spinor formalism with the only new feature being that $b_0 U^*$
is only super-Poincar\'e covariant after performing an ASPC
gauge transformation as in \pureimp. 

For example, for physical massless states, the pure spinor 
ghost number $+2$ vertex operator is 
\eqn\antif{ U^* = (\l \g^{mnpqr} \l) C_{mnpqr}(x,\t)}
where $C_{mnpqr}(x,\t)$ is a five-form superfield. $U^*$ is annihilated by
$Q$ if
$$\l^\a(\l\g^{mnpqr}\l) D_\a C_{mnpqr} =0$$
for any pure spinor $\l^\a$,
which implies that
\eqn\eomaf{\g^{mnpqr}_{(\a\b} D_{\g)} C_{mnpqr} = \g^m_{(\a\b} H_{\g)m}}
for some $H_{\g m}$. Furthermore, $\d U^* = Q \Omega$ implies the gauge
transformation
\eqn\gtaf{\d C_{mnpqr} = (\g_{mnpqr})^{\a\b}D_\a \Lambda_\b}
where $\Omega= \l^\a\Lambda_\a$.
To this author's knowledge, this is the first time that super-Maxwell
antifields have been described in $d=10$ superspace. 

To perform a component analysis of $C_{mnpqr}$, it is convenient
to choose a gauge in which the lowest component of $C_{mnpqr}$ is
cubic in $\t^\a$. In this gauge, 
\eqn\expC{C_{mnpqr} = (\t \g_{[mnp}\t)(\t \g_{qr]})^\a\psi^*_\a(x) +
(\t \g_{[mnp}\t)(\t \g_{qr]s}\t) a^{*s}(x) + ...}
where $\psi_\a^*(x)$ is the antifield for the photino $\psi^\a(x)$,
$a^{*s}(x)$ is the antifield for the photon $a_s(x)$, and
all component fields in $...$ can be expressed in terms of 
$\psi^*_\a$ and $a^{*s}$. These antifields
satisfy the equation of motion $\p_m a^{*m}=0$ with the gauge
transformations 
\eqn\gaugaf{\d a^{*m} = \p_n(\p^m \omega^n - \p^n \omega^m),\quad
\d \psi^*_\a = \g^m_{\a\b}\p_m \phi^\beta}
where $\omega^m$ and $\phi^\beta$ are gauge parameters.
As usual, the gauge transformations of the antifields are related to
the equations of motion of the corresponding fields. It is interesting
to note that, up to proportionality constants, 
\eqn\compl{\langle U^* U \rangle = \int d^{10} x (  a^{*m}(x) a_m(x) +
\psi^*_\a(x) \psi^\a(x) )}
where $U$ and $U^*$ have been gauge-fixed to the form
$$U= \l^\a A_\a(x,\t) = (\l\g^m \t) a_m(x) + (\l\g^m\t) (\t\g_m)_\a
\psi^\a(x) + ...,$$
$$U^* =(\l\g^{mnpqr}\l) C_{mnpqr}(x,\t)$$
$$= 
(\l\g^m\t)(\l\g^n\t)(\t \g_{mn})^\a\psi^*_\a(x) +
(\l\g^m\t)(\l\g^n\t)(\t\g_{mnp}\t)a^{*p}(x) + ...,$$
and
$\langle ~~ \rangle$ is defined using the zero mode prescription of 
\zeromode.

\subsec{Reparameterization invariant tree amplitude prescription}

In bosonic and RNS string theories, one can compute $N$-point
amplitudes in a worldsheet reparameterization invariant manner
by defining 
\eqn\repar{A = \langle \prod_{s=1}^{N-3}(\int d^2 z_s ~\mu_s(z_s)b(z_s)) 
U_1(y_1) ... U_N(y_N)\rangle}
where $U_r$ are unintegrated ghost-number zero vertex operators. The
integrand of \repar\ coincides with the integrand of \amplrns\ when
the $N-3$ Beltrami differentials $\mu_s$ are chosen to correspond
to $N-3$ of the $y_r$'s, and differs by a BRST-trivial operator for
other choices of the Beltrami differentials. 

The reparameterization invariant prescription of \repar\ can also be used
in the pure spinor formalism, but the normalization prescription of 
\zeromode\
cannot be directly applied since the integrand of \repar\ is not
manifestly super-Poincar\'e covariant. To define an appropriate
normalization prescription which will be denoted by $\langle ~~\rangle_{ASPC}$,
use the free-field OPE's of section 2 to write 
\eqn\repp{A = \int d^2 z_1 ... \int d^2 z_{N-3} \langle 
f(z_s,\eta_r,k_r) \rangle_{ASPC}} 
where $f(z_s,\eta_r,k_r)$ is some ghost-number +3 operator
constructed from ASPC combinations of the spacetime superfields appearing
in the external vertex operators. Since $f$ is BRST invariant, the
conjecture of \conjone\ implies that 
\eqn\conimp{f = U + Q\Omega}
where $U$ is a super-Poincar\'e covariant operator and
$\Omega$ is some ASPC operator constructed from supersymmetric
combinations of the spacetime superfields appearing in the external
vertex operators.

The normalization prescription $\langle ~~\rangle_{ASPC}$ will
be defined such that for any ASPC gauge transformation $\Omega$,
\eqn\newpres{\langle U + Q\Omega\rangle_{ASPC} = \langle U\rangle}
where $\langle ~~ \rangle$ is defined in \zeromode. In other words, to apply
the normalization prescription of $\langle ~~\rangle_{ASPC}$, one
has to first remove the non-Lorentz covariant part of the integrand
by performing an ASPC gauge transformation. The conjecture of \conjone\
implies
that this procedure is unambiguous since different choices for the
ASPC gauge transformation only change the super-Poincar\'e covariant
part by $ U \to U + Q\Lambda$ where $\Lambda$ is a pure spinor gauge
transformation. But as was shown in \ampgau, 
$\langle U+ Q\Lambda \rangle = \langle U \rangle$ when $U$ and $\Lambda$
are super-Poincar\'e covariant operators.

\newsec{Open Problems and Speculations}

In this paper, the pure spinor formalism for the superstring
was related to the RNS formalism by finding a field redefinition
and similarity transformation which maps the pure spinor BRST operator
into the sum of the RNS BRST operator and $\eta_0$ ghost. Although
this map can be used to relate vertex operators and tree amplitudes in
the two formalisms, there remain at least three open problems which need to
be resolved before claiming a proof of equivalence of the two formalisms.

Firstly, because the similarity transformation $R$ of \simrone\
involves inverse
powers of $u_{ab}$, it was necessary to conjecture in \conjone\
that there exists
an ASPC gauge choice for physical vertex operators
in which these inverse powers
of $u_{ab}$ are absent. Although evidence was presented in
support of this conjecture, an explicit proof is lacking. 
Secondly, the hermiticity definition for RNS operators
was not shown to coincide with the rather unusual hermiticity definition
for pure spinor operators which was discussed in subsection (2.2).
And thirdly, it was not yet shown how to use the ASPC $b$ ghost of \bdef\
to define pure spinor loop amplitudes which coincide with the RNS 
loop amplitude prescription.

An interesting feature of this paper is the important role of ASPC
operators, i.e. operators which transform covariantly
under all supersymmetry transformations and under
$U(5)$ Lorentz transformations.
A better description of these operators,
perhaps using a harmonic superspace, might help to resolve the above
open problems. For example, it might be useful to introduce a second pure
spinor variable $\overline\lambda^\a$ into the 
pure spinor Hilbert space where $\overline\lambda^\a$ is defined to
be the hermitian conjugate of $\l^\a$.
One could then
describe
ASPC operators as
pure spinor operators with non-trivial dependence on
$\overline\l^\a$. This would make the formalism resemble the $N=2$
twistor string considered in \ref\ton
{M. Tonin, {\it World Sheet Supersymmetric Formulations
of Green-Schwarz Superstrings}, Phys. Lett. B266 (1991) 312.}
and \ref\hetw{N. Berkovits,
{\it The Heterotic Green-Schwarz Superstring on an
$N=(2,0)$ Super-Worldsheet}, Nucl. Phys. B379 (1992) 96,
hep-th/9201004.} where $\l^\a$
and $\overline\l^\a$ are pure spinor twistor-like variables satisfying the
condition $\Pi^m = \l^\a \g^m_{\a\b}\overline\l^\b$ \ref\stvz
{D.P. Sorokin, V.I. Tkach, D.V. Volkov and A.A. Zheltukhin,
{\it From the Superparticle Siegel Symmetry to the Spinning Particle
Proper Time Supersymmetry},
Phys. Lett. B216 (1989) 302.}. It is interesting
to note that the fermionic $N=2$ superconformal 
generators in the $N=2$ twistor string formalism are $G^+=\l^\a d_\a$ and
$G^-=\overline\l^\a d_\a$. Since $\oint G^+$ can be interpreted as the BRST
charge, it might be possible to interpret $\overline \l^\a d_\a$ as the 
ASPC $b$ ghost. 

\vskip 15pt
{\bf Acknowledgements:} I would like to thank Michael Bershadsky, 
Osvaldo Chand\'{\i}a, Sergey Cherkis, Paul Howe, Yoichi Kazama,
John Schwarz, Warren
Siegel, Brenno Vallilo and Edward Witten for useful discussions,
CNPq grant 300256/94-9, 
Pronex 66.2002/1998-9,
and FAPESP grant 99/12763-0
for partial financial support, and the univerisities of Caltech, Santa
Barbara, USC, Harvard and Rutgers for their hospitality.
This research was partially conducted during the period the author was
employed by the Clay Mathematics Institute as a CMI Prize Fellow.

\listrefs

\end